\documentclass[twocolum,12pt]{iopart}

\usepackage{graphicx}
\usepackage{amssymb}
\usepackage{amsfonts}
\usepackage{bbold}
\usepackage{mathrsfs}
\usepackage{color}
\usepackage{epstopdf}

\newcommand{\BEQ}{\begin{equation}}
\newcommand{\EEQ}{\end{equation}}
\newcommand{\BEA}{\begin{eqnarray}}
\newcommand{\EEA}{\end{eqnarray}}

\newcommand{\nn}{\nonumber }

\newcommand{\vk}{{\mathbf q}}

\begin{document}
\title{Effective run-and-tumble dynamics of bacteria baths}
\date{\today}

\author{M. Paoluzzi, R. Di Leonardo, L. Angelani}

\address{CNR-IPCF, UOS Roma, Dip. Fisica, Universit\`a {\em Sapienza}, P. le A. Moro 2, I-00185,
  Rome, Italy}
\ead{matteo.paoluzzi@ipcf.cnr.it}

\begin{abstract}
{\it E. coli} bacteria swim in straight runs interrupted by sudden reorientation events called tumbles.
The resulting random walks give rise to density fluctuations that can be derived analytically in the limit of non interacting particles or equivalently of very low concentrations. However, in situations of practical interest, the concentration of bacteria is always large enough to make interactions an important factor. 
Using molecular dynamics simulations, we study the dynamic structure factor of a model bacterial bath for increasing values of densities. We show that it is possible to reproduce the dynamics of density fluctuations in the system using a free run-and-tumble model with effective fitting parameters. We discuss the dependence of these parameters, e.g., the tumbling rate, tumbling time and self-propulsion velocity, on the density of the bath. 
\end{abstract}

\maketitle

%%%%%%%%%%%%%%%%%%%%%%%%%%%%%%%%%%%%%%%%%%%%%%%%%%%%%%%
\section{Introduction}
Suspensions of self-propelling particles display a complex dynamical behavior that is hard to model analytically especially in the presence of interactions.
Many different approaches have been proposed.
Starting with the seminal work of Vicsek \cite{Vicsek95}, rule-based models describe interactions between neighbors as a set of phenomenological prescribed rules that tend to favour alignment of nearby particle velocities. Those minimal models are particularly suited for numerical investigation and exhibit non-equilibrium transitions between disordered and ordered flocking state \cite{Chate08,Chate08b}.
Continuum and coarse-grained models aim at describing long wavelength behavior of active matter starting from an analogy with continuum liquid crystals models with the added ingredient of self-propulsion \cite{Baskaran08}. 
The parameters in the hydrodynamic equations can be obtained starting from different microscopical models\cite{Baskaran08,Baskaran09,Bertin06,Bertin09,Farrell12} or phenomenologically \cite{Csahok02,Toner95,Toner98}
Another possible approach is provided by the so called Òrun and tumbleÓ models \cite{Schnitzer93,Cates12}.
These models are based on the schematization of single bacterium dynamics as a sequence of linear runs interrupted by random ---Poissonian distributed--- tumbling events. The resulting movement is a random walk with a persistence length that marks the crossover between a ballistic regime at short scales and a Brownian, diffusive regime at longer scales.
The free particle dynamics is parametrized by the swimming velocity, the tumbling rate and the duration of a tumble event. These parameters can be measured either directly, through cell tracking techniques,\cite{Berg04,Berg72,Grossart01,Schneider74,Othmer88}  or indirectly, through the intermediate scattering function (ISF) as obtained from dynamic light scattering or more recently from differential dynamic microscopy\cite{Martens12,Wilson11,Martinez12,Cerbino08,Boon74,Stock78,Giavazzi09}.
Experimental ISF are generally fitted using analytic results obtained for the non-interacting case. Moreover, the high variability of motility characteristics that is encountered in real bacterial populations, makes experimental ISFs practically insensitive to the details of single particle trajectories like tumbling rate or tumble duration. Therefore experimental data are usually fitted with a free (non-interacting) and non-tumbling theoretical ISF convoluted with the unknown distribution of swimming speeds. The retrieved fitting parameters often show a q-dependence in single particle quantities like the swimming speed, that is usually ascribed to imaging artifacts or the effect of tumbling rate \cite{Martinez12}. 
However, the dynamic structure factor for an ensemble of non interacting run and tumble particles has an analytic expression even in the presence of tumbling and of a finite tumbling time \cite{Martens12, Angelani13}. We refer to these last results as the {\itshape free theory}, corresponding to the case of
non-interacting bacteria, e. g., a gas of {\itshape E. Coli}.
%In this paper we use the expression obtained in \cite{Angelani13} as a model for
%the dynamic structure factor in interacting bacterial suspensions. 
%We perform molecular dynamic simulations of an active system choosing motility and mobility parameters to mimic
%the behavior of {\itshape E. Coli} bacterial bath. 
The main aim of this paper is to discuss
the applicability of a free theory to describe the properties of an ensemble of interacting
run-and-tumble particles in a wide range of densities.
Studying the dependence on density of the effective parameters of
the theory we map the interacting dynamics into a non-interacting system with
effective values of tumbling rate, run velocity and tumbling time (Cfr. Fig(\ref{fig:trac})). 

In section \ref{MDS} we introduce the numerical model, the observables
we are interested and their theoretical expressions obtained with the
free run-and-tumble model. In section \ref{results} we report the density
and wave vector-dependence of the effective parameters and discuss the
results. We also introduce an effective diffusivity and discuss the possible existence
of a spinodal decomposition in the active bath \cite{Tailleur08}.

\section{Molecular Dynamics simulations}\label{MDS}
In order to study the effects of interactions on the motility of single cells, 
we have performed Molecular Dynamics simulation 
of a run-and-tumble model \cite{Angelani09,Angelani10,Angelani11}.
The simulations involve $N$ elongated hard body cells in two dimensions
closed in a square box of side $L$ with periodic boundary conditions.
Each cell ---labelled by $i=1,\dots,N$--- is represented by the sum of $p$ short-range repulsive potential 
centered at equality spaced locations $\delta^\beta$, where $\beta$ runs from $1$ to $p$. The center of mass of $i-$th cells, will be
indicated by the vector $\mathbf{r}_i$, the $\beta-$th center by $\mathbf{r}_i^\beta$
and the orientation of the $p-$bodies with $\mathbf{\hat{e}}_i$
\BEA
\mathbf{r}_i^\beta &=& \mathbf{r}_i + \delta^\beta \mathbf{\hat{e}}_i \\ \nn
\delta^\beta &=& \frac{(l - a)(2 \beta - p - 1)}{2 p - 2}\, .
\EEA
In such a picture the cell body is modeled as a 
chain of spheres (disks) rigidly connected, mimic a
prolate spheroid of aspect
ratio $\alpha=a/l$, where $l$ is the length and $a$ the thickness of the cell.
We have specialized our simulations to the case $p=2$, i. e., the 
active object is an elongated spheroid with two centers and aspect
ratio $\alpha=\frac{1}{2}$. 
At low Reynolds numbers \cite{Purcell77} equations of motion for the center of mass and angular velocities read \cite{Kim05}
\BEA\label{eqm}
\dot{\mathbf{r}}_i &=& \mathbf{M}_i \cdot \mathbf{F}_i \\ \nn
\dot{\mathbf{\theta}}_i &=&\mathbf{K}_i \cdot \mathbf{T}_i 
\EEA
where $\mathbf{M}_i$ and $\mathbf{K}_i$ are the translational 
and rotational mobility matrices of the $i-$th swimmer
\BEA
\mathbf{M}_i &=& m_{\parallel} \mathbf{e}_i \otimes \mathbf{e}_i  + m_{\perp} \left( \mathbb{1} -  \mathbf{e}_i \otimes \mathbf{e}_i \right)   \\ \nn
\mathbf{K}_i  &=& k_{\parallel} \mathbf{e}_i \otimes \mathbf{e}_i  + k_{\perp} \left( \mathbb{1} - \mathbf{e}_i \otimes \mathbf{e}_i \right) \, ,
\EEA
$ \mathbf{F}_i$ and $\mathbf{T}_i$ are the total force and the
total torque acting on the swimmer
\BEA
 \mathbf{F}_i  &=& f_0 \mathbf{\hat{e}}_i(1 - \sigma_i) + \sum_{j\neq i} \mathbf{f}(\mathbf{r}_i^\alpha - \mathbf{r}_j^\beta) \\ \nn
 \mathbf{T}_i  &=& \mathbf{t}_r \sigma_i + \mathbf{\hat{e}}_i \times \sum_{j\neq i} \delta^\beta \mathbf{f}(\mathbf{r}_i^\alpha - \mathbf{r}_j^\beta) \, .
\EEA
Performing two dimensional simulations, $k_{\parallel}$ does not
play any role.
To describe swimmers like {\itshape E. Coli} we choose
$k_{\perp}=4.8$, $m_\parallel=1$ and $m_\perp=0.87$ \cite{Angelani09}.
In this paper all the quantities are expressed in internal unit $l=m_\parallel=f_0=1$.
Reasonable values for {\itshape E. Coli} bath are, $l\simeq 3 \,\mu$m, $m_\parallel \simeq 60 \,\mu$ms$^{-1}$ pN$^{-1}$,
$f_0\simeq 0.5 \,$pN. Realistic values for time spending in {\itshape tumble state} and
{\itshape running state} are, respectively, $0.1$ s and $1$ s.

Self-propulsion is modeled by
the presence of a state variable
$\sigma_i$ which can assume the 
values $1$ and $0$. 
During the {\itshape running state} $\sigma_i = 0$ and
the object, due to the active force $f_0$, is self-propelled along the direction $\mathbf{\hat{e}}_i$
with constant velocity $v=m_\parallel f_0$. Stochastically the runners randomize
the direction of motion and the state variable changes discontinuously
assuming value $1$. This contribution is given by 
the term $\mathbf{t}_r$ which is a random torque that randomizes the 
swimming direction $\mathbf{\hat{e}}_i$.
The pair force $\mathbf{f}(\mathbf{r})$, describing cell-cell interaction, is chosen purely repulsive
$
\mathbf{f}(\mathbf{r})=\frac{A \mathbf{r}}{\mathbf{r}^{n+2}}
$
and the coefficient $A$ is fixed such that two swimmers facing
head to head on the same line would be in equilibrium at the distance $a$
(the thickness of the hard body)
$
A = f_0 a^{n+1}\, 
$
(for the soft-sphere repulsion we choose $n=12$).
Eqs. (\ref{eqm}) are numerically integrated for
$4 \times 10^{5}$ steps through Runge-Kutta 
second order scheme \cite{Press92} with a time step of $10^{-3}$. 
The simulations are performed at several values
of density, from $\rho=0$ (non interacting bacteria) 
to $\rho=0.7$, varying the number of bacteria in a box of side $L=70$
(we consider a fixed box length in order to have
the same wave vectors at each density) from
$N=576$ ($\rho=0.1$) to $N=3481$ ($\rho=0.7$). Averages 
 over $N_s(\rho)$ samples are considered, with $N_s\in[30,10]$.
In the range of investigated densities the system always remains homogeneous and displays no clear phase separations.  This is a fundamental requirement when trying to describe the dynamics of density fluctuations using an effective free theory. 

\subsection{Observables}
We choose to describe density fluctuations in our system through the Intermediate Scattering Function (ISF) and  and its {\itshape frequency-domain} Fourier transform: the Dynamic Structure Factor (DSF). The two functions can be accessed directly from experiments \cite{Boon74,Berne00} and are particularly suited for analytical calculations \cite{Martens12,Angelani13}.
In particular, the ISF is the time correlation function of the spatial Fourier transform of density fluctuations:
\BEA \label{inv}
F(\mathbf{q},t)&=&\frac{1}{N}\frac{\langle \hat{\rho} (\mathbf{q},t+t^\prime) \hat{\rho} (-\mathbf{q},t^\prime)\rangle_{t^\prime}}{\langle \hat{\rho} (\mathbf{q},t^\prime) \hat{\rho} (-\mathbf{q},t^\prime)\rangle_{t^\prime}} \\ \nn
\hat{\rho} (\mathbf{q},t)&=&\int d\mathbf{r} \, e^{-i \mathbf{q} \cdot \mathbf{r}}\rho(\mathbf{r},t) \\ \nn
\rho(\mathbf{r},t)&=& \sum_i \delta\left( \mathbf{r} - \mathbf{r}_i(t)\right) \\ \nn
\langle \mathcal{O}(t) \rangle_t &\equiv& \frac{1}{T}\int_0^T dt \, \mathcal{O}(t) \, .
\EEA
Since the numerical simulations are performed in a square box of side $L$ with
periodic boundary conditions, our study in the reciprocal space is limited by the minimum wave-length
$q_m=\frac{2 \pi}{L}$ and, consequently, we have computed ISF for wave vector
of the form
\BEQ
\mathbf{q}=q_m \mathbf{n}, \;\; \mathbf{n}=(n_1,n_2)
\EEQ 
with $n_i\in\mathbb{Z}$ for $i=1,2$. run-and-tumble dynamics is isotropic: we can average over the modulus $q=| \mathbf{q}|$ and, in order to improve the
statistic, we also average $F(\mathbf{q},t)$ over moduli $q_k=| \mathbf{q_k}|$ 
close to $q$. Choosing an opportune $\epsilon$ one has 
\BEQ
F(q,t)=\frac{\sum_{| q - q_k | <  \epsilon }F(q_k,t)}{\Omega(q)}\, ,
\EEQ
where $\Omega(q)$ is a normalization factor.
\begin{figure*}[t!]
%\hspace*{-.5in}
\hspace*{.5in}
\centering
\resizebox{.9\columnwidth}{!}{%
\includegraphics[width=0.6\columnwidth]{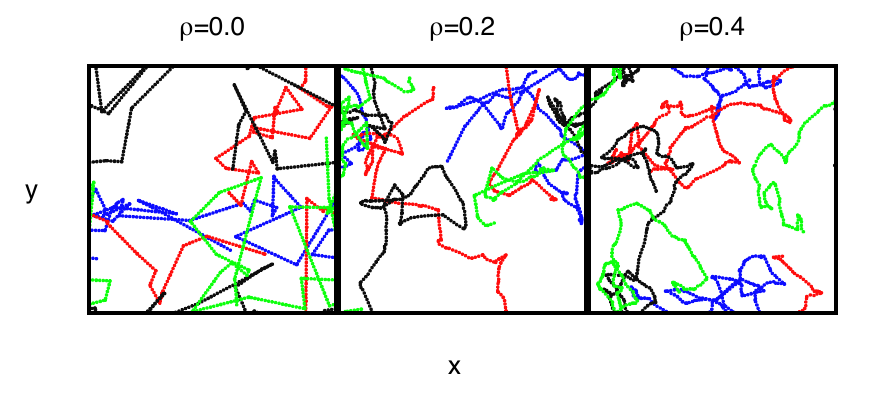}
}
\centering
\caption{Trajectories of run-and-tumble bacteria from non interacting to interacting case. Repulsive
{\itshape soft-core} interaction, increasing density, smooths trajectories
on large scale. In this work we try to describe a smooth (interacting) trajectories 
with a non-interacting one through an opportune choice of the
active parameters.}
\label{fig:trac}       % Give a unique label
\end{figure*}
\begin{figure}[t!]
\centering
\resizebox{.85\columnwidth}{!}{%
\includegraphics[width=1.\columnwidth]{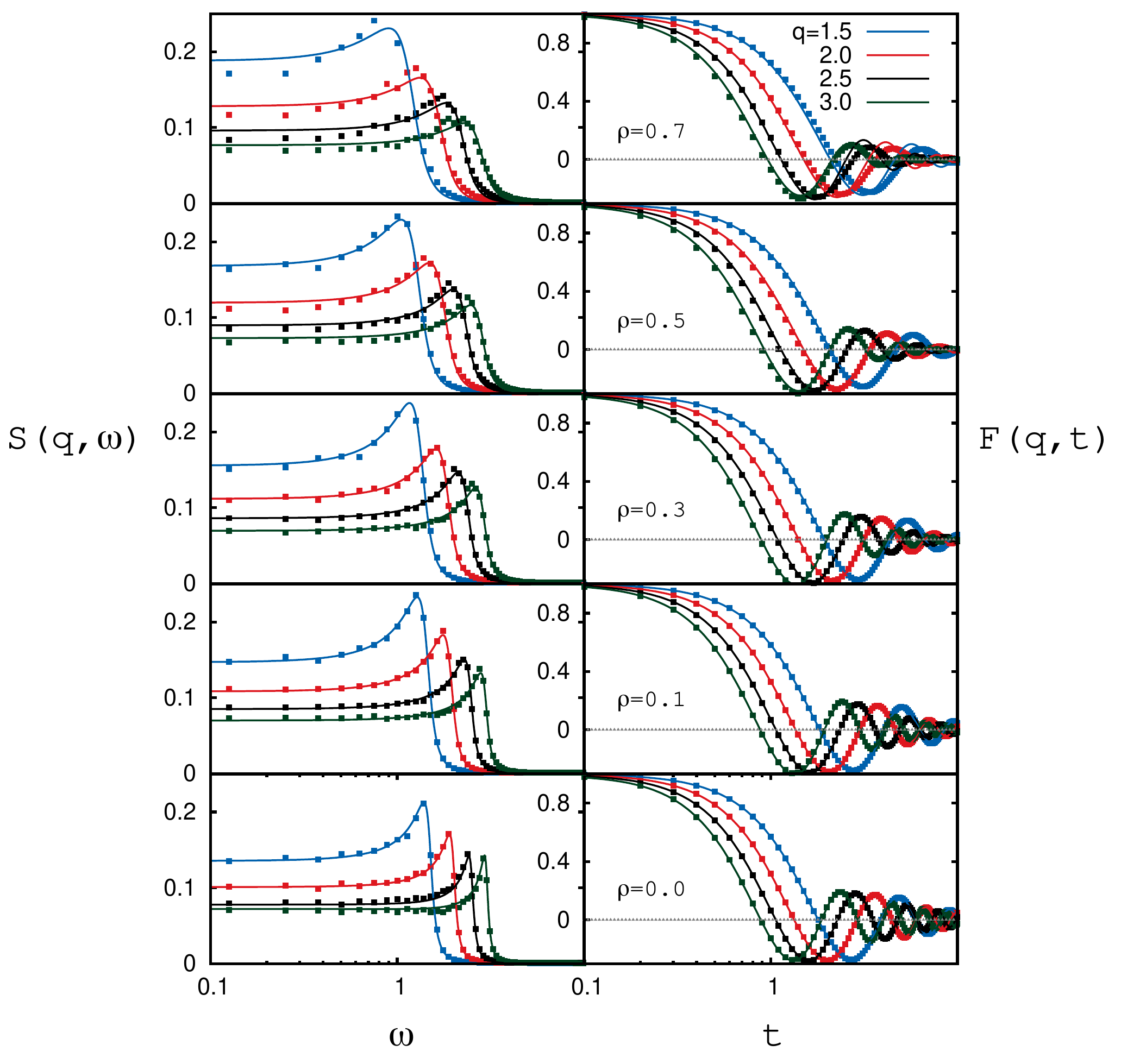}}
\caption{Left panel: comparison between Dynamic Structure Factor for a molecular dynamics 
simulation of an active interacting bath in two dimensions at $\rho=0.0,\dots,0.4$ (symbols) and theory (eq. (\ref{luca})-lines) 
fitted with effective values of $\lambda$, $v$ and $\tau$. 
Right panel: comparison between Intermediate Scattering Function computed by eq. (\ref{inv}) for the values of effective parameters (lines)
and directly measured by simulations data (symbols). For $\rho=0.0$, where the theory is exact, the lines are the analytic solution.
At the largest shown density ($\rho=0.7$) the free theory is unable to describe the simulated correlation functions. }
\label{fig:sqwfqt}       % Give a unique label
\end{figure}
DSF, numerically obtained by Fast Fourier Transform of ISF, is normalized with respect to the area
\BEQ
S(\mathbf{q},\omega)=\frac{\Re \left[ \tilde{S}(\mathbf{q},\omega) \right]}{\int_{0}^{\infty} d\omega\, \Re \left[ \tilde{S}(\mathbf{q},\omega) \right]} \\ \nn
\EEQ
where
\BEQ
\tilde{S}(\mathbf{q},\omega)=\frac{1}{2 \pi}\int_{-\infty}^{+\infty}dt\, e^{-i \omega t} F(\mathbf{q},t) \, .
\EEQ
To obtain informations about the motility properties
of an active bath, we need of a model for the ISF (or, equivalently, for the DSF) 
%
%%%%%%%%%%%%%%%%%%%%%%%%%%%%%%%%%%%%
%%%%%%%%   LUCA   %%%%%%%%%%%%%%%%%%
%%%%%%%%%%%%%%%%%%%%%%%%%%%%%%%%%%%%
A theoretical expression for the Laplace-Fourier transform of the probability distribution
of free run-and-tumble particles can be obtained within the continuous-time random walk approach \cite{Klafter11}. 
By considering a random walk consisting of two independent alternating phases 
(run at constant speed interrupted by tumble events) one can write the 
probability distribution function as a sum of all the possible convolution products of the 
propagators of the two phases (see Ref. \cite{Angelani13} for details).
For the case considered here, with a Poissonian distribution of tumble events (with time rate $\lambda$)
and a constant time duration of the tumble phase ($\tau$), we obtain the following expressions:
\BEA \label{luca} \nn
P({\bf q},z,\tau)&=&\frac{1}{1 + \lambda \tau} \left( \mathscr{F}({\bf q},z,\tau) G({\bf q},z,\tau)
+ g(z,\tau) \right)\\ \label{pkz}
\mathscr{F}({\bf q},z,\tau)&\equiv& \frac{F_0({\bf q},z)}{1 - \lambda e^{-z \tau}F_0({\bf q},z)}\\ \nn
G({\bf q},z,\tau)&\equiv& \left[ 1+\frac{\lambda}{z}\left( 1 - e^{-z \tau} \right) \right]^2 \\ \nn
g(z,\tau)&\equiv&\frac{\lambda \tau}{z}\left\{ 1 - \frac{1 - e^{-z\tau}}{z \tau} \right\} \,,
\EEA 
where $F_0({\bf q},z)$ is the Laplace-Fourier transform of the ballistic propagator of the theory
(see Eq. (17) of Ref. \cite{Angelani13}, having chosen a
constant duration $\tau$ of tumble event).
In two dimensions one has \cite{Martens12,Angelani13}
%%%%%%%%%%%%%%%%%%%%%%%%%%%%%%%%%%%%%%
\BEQ
F_0({\bf q},z)=\frac{1}{\sqrt{\left(z + \lambda \right)^2 + \left( q v \right)^2}} \; 
\EEQ 
with $v$ velocity of the ballistic run.
$P({\bf q},z,\tau)$ is related to the
DSF through the relation
\BEQ
\tilde{S}(\vk,\omega,\tau)=\lim_{\delta\to 0}P({\bf q},i \omega +  \delta,\tau)\, .
\EEQ
Eq. (\ref{luca}) can be written in time domain (details can be found in the Appendix)
\BEA \nn
P({\bf q},t,\tau)&=&\frac{1}{1 + \lambda \tau}\left[ \int_0^t \, dt^\prime \, \mathscr{F}({\bf q},t^\prime,\tau) G(t-s)+ g(t,\tau)\right] \\ 
g(t,\tau)&\equiv& \lambda (t-\tau) (\theta(t-\tau) - \theta(t) ) \\ \nn
\mathscr{F}({\bf q},t,\tau)&\equiv&\mathcal{L}^{-1}\left[ \frac{F_0({\bf q},z)}{1 - \lambda e^{-z \tau}F_0({\bf q},z)} \right] \\ \nn
G(t)&=&\delta(t)+\lambda [(2 + t \lambda)\theta(t) + \lambda(t-2\tau)\theta(t-2\tau) \\ \nn
&-& 2(1+t\lambda -\lambda \tau)\theta(t-\tau)]
\EEA
operator $\mathcal{L}^{-1}\left[ \dots \right]$ is the inverse Laplace-transform.
For the ISF one has $P({\bf q},t,\tau)=F({\bf q},t)$. %
In the bottom of Fig. (\ref{fig:sqwfqt}) we report the comparison
between theory (Eq. (\ref{luca})) and simulations for a non-interacting ($\rho=0$) bacterial bath.
\section{Results}\label{results}
At each density from $\rho=0.1$ to $0.7$, we fit the DSF with
the free theory (Eq. (\ref{luca})) obtaining effective 
values for the parameters $\lambda$, $v$ and $\tau$ (see Fig. (\ref{fig:sqwfqt})).
Within this framework we can map the trajectories of interacting 
bacteria in the trajectory of a free bacterium with 
effective motility parameters set to $v_{eff}$, $\lambda_{eff}$ and $\tau_{eff}$.  
In Fig. (\ref{fig:trac}) we show some trajectories as a function
of density. Our aim is to describe the trajectories at $\rho\neq0$ through 
a non interacting run and tumble.
%

\begin{figure}[h!]
\centering
\resizebox{0.5\columnwidth}{!}{%
\includegraphics[width=0.8\columnwidth]{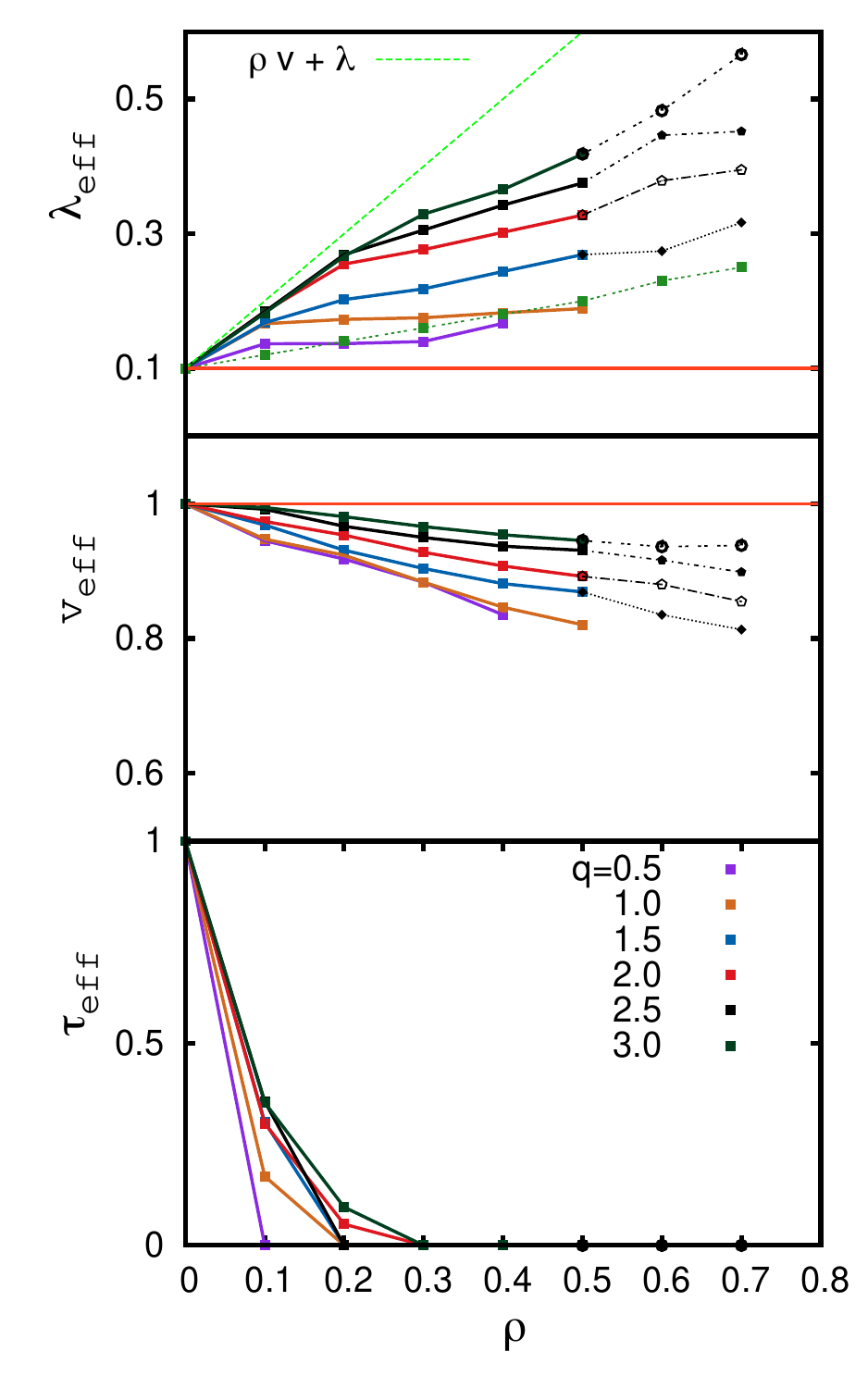}}
\caption{Effective parameters as a function of density. 
From top to bottom, tumbling rate, ballistic velocity and
tumbling time. Green points in the top panel are
effective tumbling rate computed by Eq. (\ref{msd}). Green line is Eq. (\ref{path}). Increasing
density, collision events at small $q-$values ($\rho>0.4$ and $q=0.5,1.0$) 
do not allow to obtain reasonable parameters. Black dashed lines are the effective parameters for $\rho>0.4$,
region of the phase diagram where the quality of the fit of $S(q,\omega)$ degrades (see Fig. (\ref{fig:sqwfqt})). 
As is shown, effective tumbling rate measured through DSF tends to
the effective tumbling rate obtained from MSD in the limit $q \to 0$.
%Results obtained by mean-square displacement tends to the $q \to 0$ limit
%of the results obtained by DSF.
}
\label{fig:para}       % Give a unique label
\end{figure}

%We can not perform a mapping at each density and at each $q-$values:
%the interaction introduces a collision rate that grows with the density of the systems.
In Fig. (\ref{fig:sqwfqt}) we show the results for different densities and wave vectors 
$q=1.5,2.0,2.5,3.0$. At the largest investigated density the free theory is unable to reproduce the simulated DSF and the discrepancies are larger at lower $q$ values. This may indicate the onset dynamical correlations that cannot be captured by an effective free theory.
Fig. (\ref{fig:para}) shows the $q-$dependence of the parameters. Increasing
density, collision events at small $q-$values ($\rho>0.4$ and $q=0.5,1.0$) do not allow to obtain reasonable parameters.
At each density one has $\lambda_{eff}(q)>\lambda$, $v_{eff}(q)<v$ and $\tau>\tau_{eff}(q)$.
Increasing density, $\tau_{eff}(q)\to 0$ at each $q$. 
The observed increasing of tumbling rate $\lambda_{eff}(q)$, by increasing density, indicates that, due to collisions, the trajectory  of a single {\itshape effective-free} bacterium, observed on a spatial scale of order $1/q$ is more broken than in the non-interacting  case (Cfr. Fig. (\ref{fig:trac})). As we can see, $\lambda_{eff}(q)$ increases with $q$ and approaches a limiting value. 
%This is due to the competition between self-propulsion and reorientation of velocity due to short-range interactions. 
Roughly speaking, at low densities, we can think of $\lambda_{eff}$ as a sum of two contributions: 
tumbling rate $\lambda$ and collision rate $\lambda_{int}$
\BEA\label{path}
\lambda_{eff} (\rho)&=& \lambda + \lambda_{int} \\ \nn
\lambda_{int}&\sim& \gamma \rho v + o(\rho^2) \, 
\EEA  
where $\gamma$ is the characteristic particle size.
In the top of Fig. (\ref{fig:para}) we show the behavior of $\lambda_{eff}(\rho)$ together with Eq. (\ref{path}). 
Studying the behaviour of effective velocity, one has $v_{eff}(q)<v$ and $v_{eff}(q)\to v$ increasing $q$:
this correspond to the purely ballistic scale of the theory.
%This result is in accord with DDM analysis \cite{Wilson11} where a dependence
%of free-swim velocity on wave vector with $v(q)<v$ is observed.  
%
%Effective velocity depends  on the wave number and varies both in space
%and in the density.
%
We recall that, in numerical simulations of run-and-tumble model with steric interaction, the
self propelled velocity $v$ is fixed during the simulations. It is well known in literature
how self-propulsion, in the hydrodynamical limit, changes the effective diffusivity of self-propelled
roads \cite{Baskaran08}. On the other hand, at mean field level, to allow a self-trap phenomenon
we have to chose an opportune dependence of swim velocity on the density \cite{Tailleur08}.
In particular it holds for a modified version of the Vicsek model discussed in \cite{Farrell12}, where
an exponentially decrease of the velocity with local density is considered. In our simulations,
the excluded volume interaction renormalizes the self-propulsion velocity and the 
renormalization is $q$-dependent.
The mean-free path due to the self-propulsion is given by $v/\lambda $: in term of
effective parameters, $\lambda_{eff}$ increases while $v_{eff}$ decreases so that
the mean free-path is shorter than in free case.
Finally $\tau_{eff}$ decreases due to running bacteria pushing on tumbling ones.
The $q-$dependence underlines a inhomogeneity in space of the effective
parameters: we can not fix only one values for $\lambda$, $v$ and $\tau$
to build an effective free theory. Nevertheless, fixing the scale, i. e., a degree
of resolution, we can describe the trajectories of the interacting system with
a non interacting one.

\begin{figure}[t!]
%\centering
\resizebox{0.33\columnwidth}{!}{%
\includegraphics[width=0.3\columnwidth]{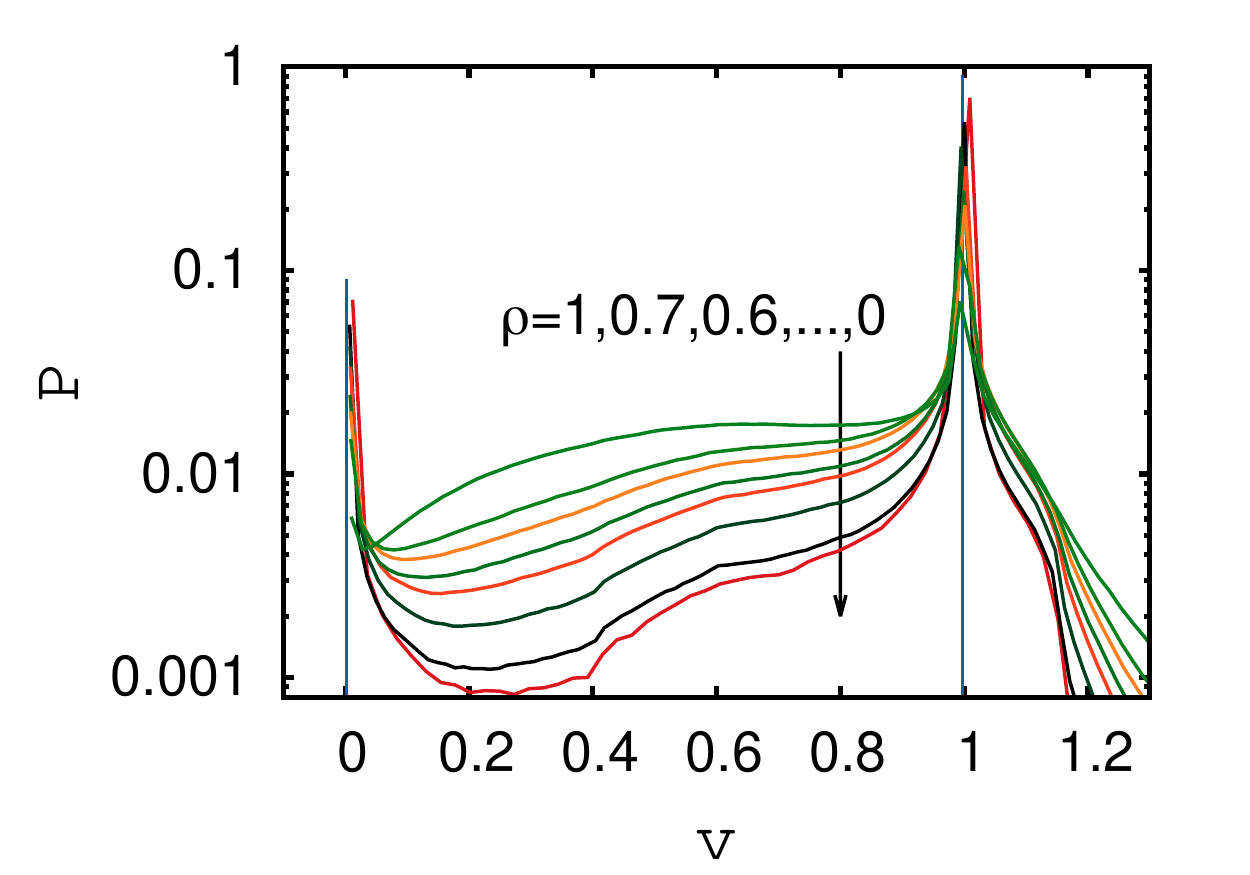}}
\resizebox{0.33\columnwidth}{!}{%
\includegraphics[width=0.3\columnwidth]{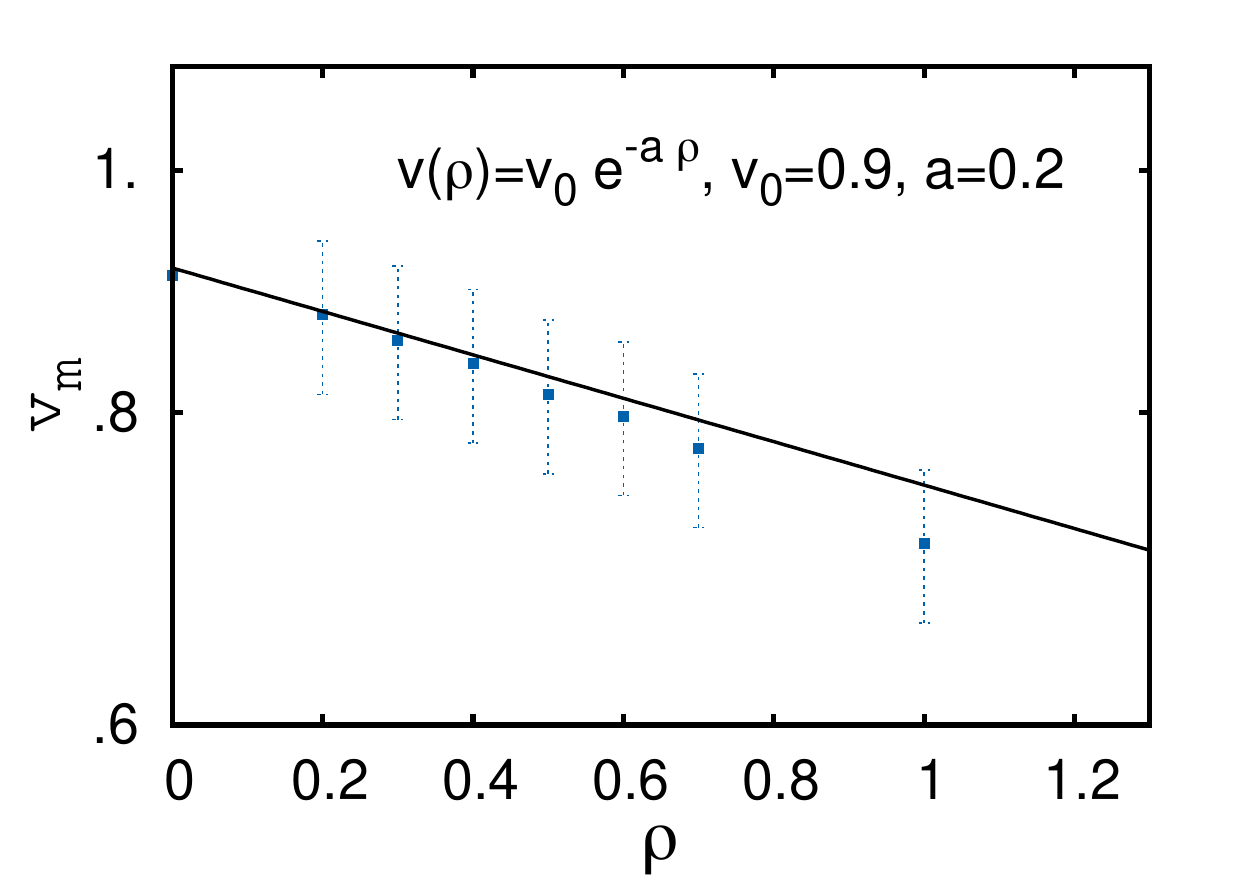}}
\resizebox{0.33\columnwidth}{!}{%
\includegraphics[width=0.3\columnwidth]{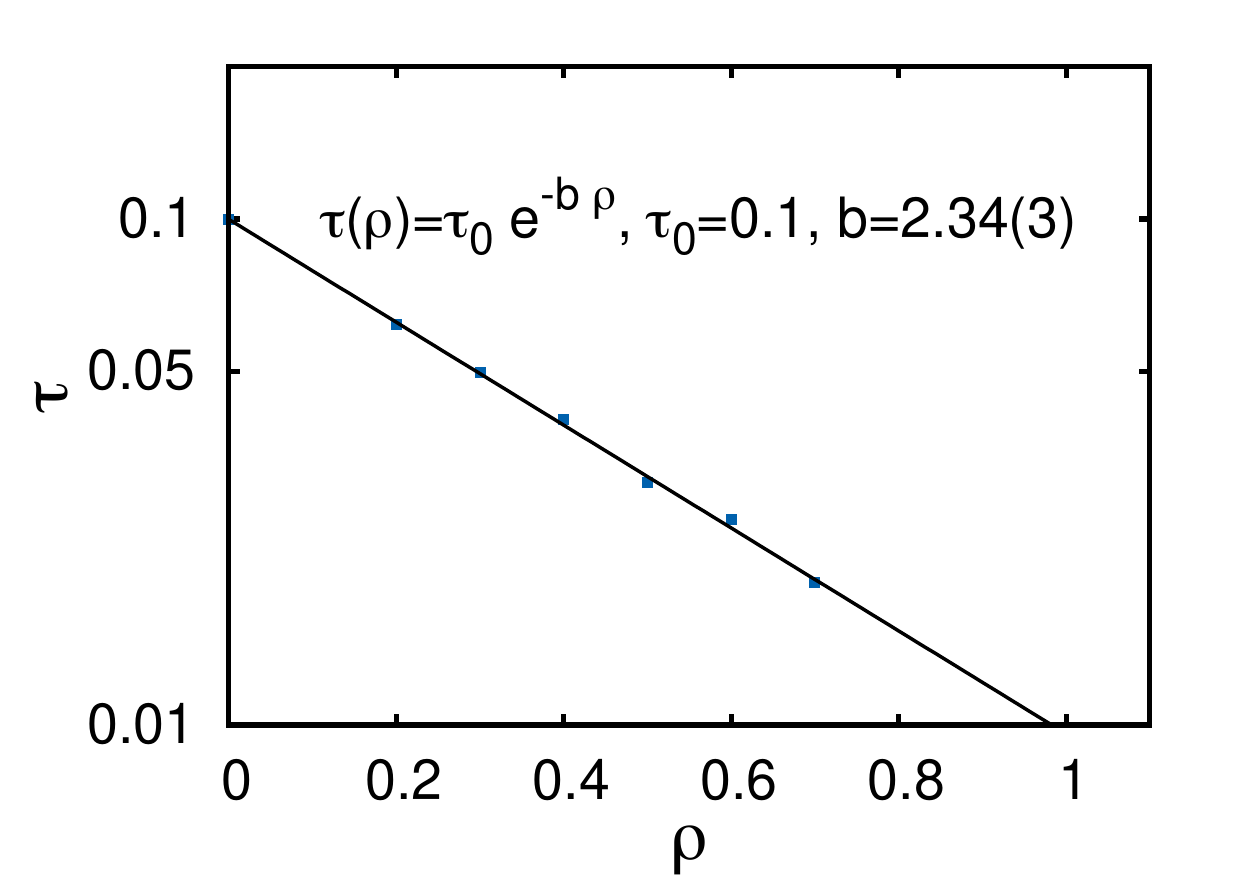}}
\caption{From left to right 
probability distribution of velocity $P(v)$,  dependence on density of average velocity and time averaged
spends in tumbling state $\tau_m$ obtained by $P(v)$.Data are for densities $\rho=0.0,0.1,0.2,0.3,0.4,0.5,0.6,0.7,1.0$.}
\label{fig:vel}       % Give a unique label
\end{figure}

Self-propelled models display a crossover
between ballistic and diffusive regime. The time scale where this happens is fixed
by the tumbling rate.
%While the behavior of the ballistic velocity gives information about
%the trajectories of a single bacterium, in order to study the collective properties
%of the system, we have to look at observables that describe the bath on large scale.
%In particular our results at fixed $q$ must be consistent in the limit of $q\to 0$. 
For free $RT$ models, mean-square displacement reads \cite{Angelani13}
\BEA\label{msd}
r^2_{MSD}(t)&=&\frac{4 D}{\lambda}\left( \lambda t - 1 + e^{-\lambda t}\right) \\ \nn
D&=&\frac{\frac{v^2}{2 \lambda}}{1 + \lambda \tau} \, ,
\EEA
being $D$ diffusivity.
Using Eq. (\ref{msd}) one can compute the crossover time $t_{cross}$ between ballistic
and diffusive regime
\BEQ\label{cross}
t_{cross}(\lambda)=\frac{2}{\lambda}.
\EEQ 
%The definition of an effective tumbling rate must be consistent with
%a $t_{cross}(\lambda_{eff})$. In the right panel of Fig. (\ref{fig:msd}) we show
%the crossover predicted by Eq. (\ref{cross}) according to the value of $\lambda_{eff}$.
On the other hand, from our analysis in $q-$domain we can define a scale-dependent diffusion coefficient 
\BEQ\label{deff}
D_{eff}(q)\equiv \frac{v_{eff}(q)^2}{2 \lambda_{eff}(q)} \frac{1}{1 + \lambda_{eff}(q) \tau_{eff}(q)}\, .
\EEQ 
We find that $D_{eff}(q)<D$ with $D_{eff}(q)$ approaching $D$ for short wave vectors.
Since $D_{eff}(q)$ is $q-$dependent, it can be interpreted like
an inhomogeneous diffusion coefficient.
Because the effective parameters vary on both $q$ and $\rho$, in order to compare the results in $q-$space with other observables, we study
the shape of the probability distribution of velocities $P(v)$, the mean velocity $v_m$, the average
time spent in the tumbling state $\tau_m$ (Fig. (\ref{fig:vel})). 
The mean velocity is obtained from $P(v)$
\BEQ
v_m=\int_0^{v_{max}}  dv\,P(v) v \, .
\EEQ
We study the probability distribution of velocities for densities $\rho=0.1,0.2,...,0.7,1.0$. 
At high density we can not think to the bacterial bath as a non 
interacting ensemble of bacteria: it is not allow to fit DSF through Eq. (\ref{luca}).

\begin{figure}[t!]
\centering
\includegraphics[width=0.5\columnwidth]{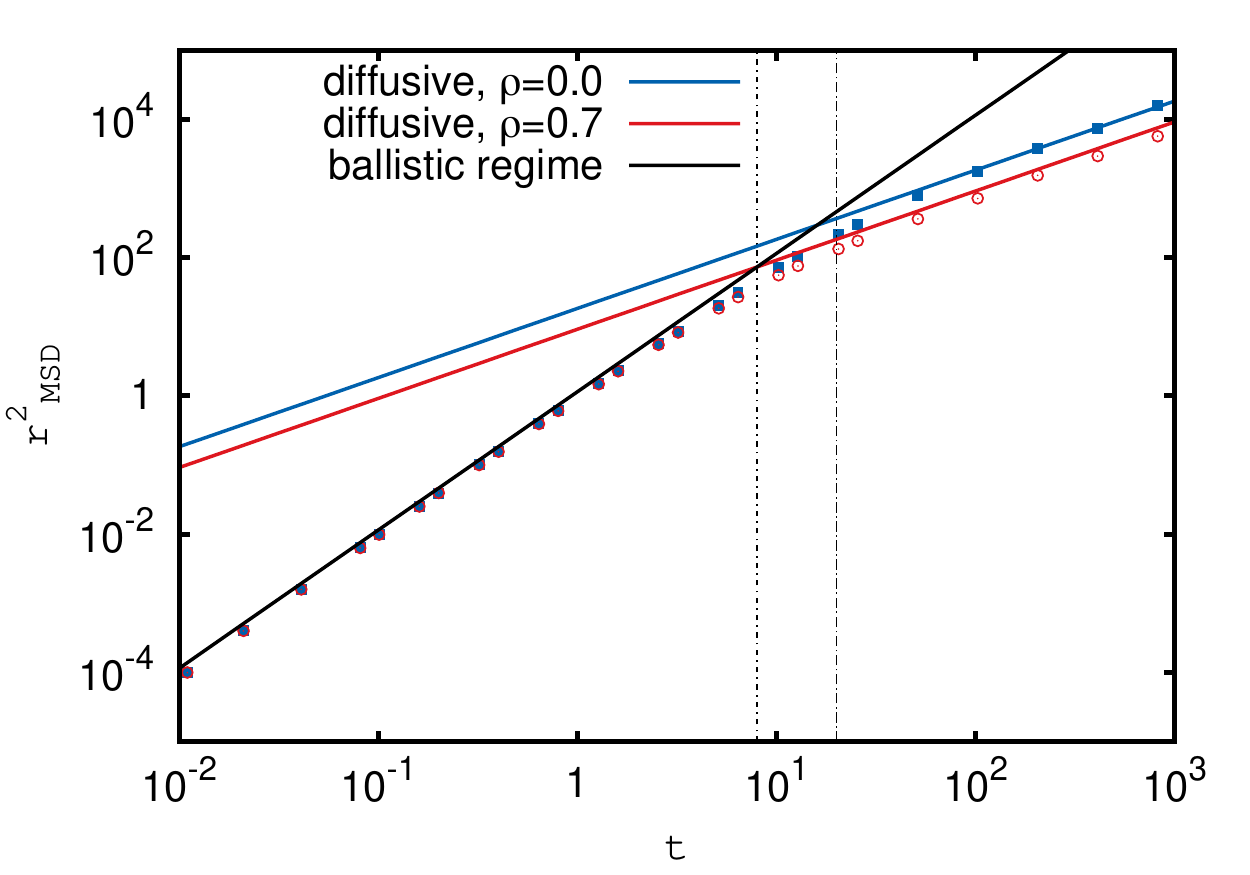}
\caption{Mean-square displacement for a non-interacting bath ($\rho=0$) and interacting ($\rho=0.7$).
Solid lines are, respectively, theory for run-and-tumble dynamics (Eq. (\ref{msd})) and effective-theory 
fitted by the same equation with $D$ and $\lambda$ as free parameters. Dashed lines are
the crossover time, compute through Eq.(\ref{cross}), from ballistic to diffusive regime. Increasing density
the crossover time decreases.}
\label{fig:msd}       % Give a unique label
\end{figure}

\begin{figure}[t!]
\centering
\includegraphics[width=0.5\columnwidth]{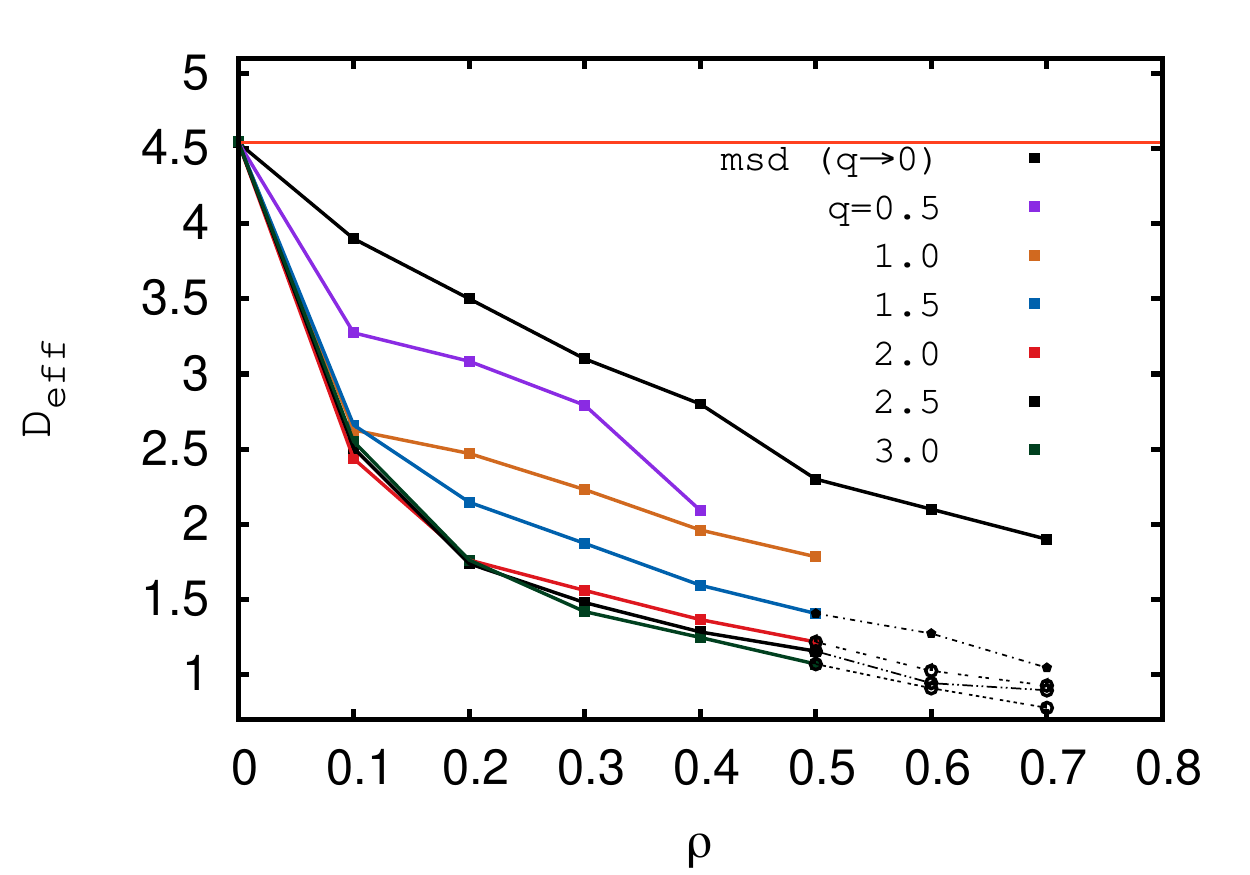}
\caption{ 
Effective diffusivity on $q-$scale (from Eq. (\ref{deff})) and diffusivity computed by mean-square displacement (black) which 
corresponds to the limit $q \to 0$ of $D_{eff}(q)$. Straight red line is the diffusivity at $\rho=0.0$.
Black dashed lines are effective diffusivity at high density.}
\label{fig:diff}       % Give a unique label
\end{figure}

In Fig. (\ref{fig:msd}) we show the mean-square displacement for non-interacting and interacting bacteria, from
which a value of $D$ can be obtained fitting the data with Eq. (\ref{msd}). A comparison between $D$ and $D_{eff}(q)$
is in Fig. (\ref{fig:diff}). As it can see, the effective diffusivity approaches the long-scale diffusivity
as $q\to 0$.
The dependence on density of $v_{m}$ and $\tau_m$ can be checked looking at $P(v)$.
Due to the presence of tumbling, the shape of $P(v)$ at $\rho=0$ is bimodal but,
increasing density, assumes a continuum support different from zero between $0$ and $v_{max}$: it is consistent
with $\tau_{eff}\to 0$. 
Mean velocity is a decreasing function of density and the functional dependence assumes an exponential form $v_m(\rho)=v_0 e^{-a \rho}$, whit $v_0=0.914(3)$ and
$a=0.200(5)$. 
From the knowledge of $P(v)$, we can define the {\itshape average time} $\tau_m$ spent
in tumbling state as the probability to have $v=0$
\BEQ
\tau_m \equiv \int_0^{v_{max}}  dv\,P(v) \delta\left( v \right) = P(0)\, .
\EEQ
The life time of the tumbling state 
decays exponentially in density following the expression $\tau_m(\rho)=\tau_0 e^{-b \rho}$ with $\tau_0=0.1$, equal to 
the time spent in tumbling state when the bacteria do not interact,  and $b=2.34(3)$ (cfr. Fig. (\ref{fig:vel})).

Recently, the possibility to have a spinodal decomposition of a run-and-tumble bath has been related to the existence of a fast decreasing velocity as a function of density, the
condition being $dv(\rho) / d\rho<-v / \rho$ \cite{Tailleur08}. We show that the presence of steric interactions results in a density dependence of velocity ($v_m$ or $v_{eff}(q)$) that is too weak to satisfy that condition. This finding is consistent with the fact that we never observe a spinodal decomposition in our simulations. A larger density effect on velocity may arise from hydrodynamic effects, biochemical interactions or other mechanisms that are not included in the present model.
\section{Conclusions}
We have performed a molecular dynamics study of interacting {\itshape run and tumble} bacteria with short-range steric repulsion in two dimensions.
Motivated by the knowledge of the analytic expression for the DSF of the free theory,
we have performed a systematic study in density of $S(q,\omega)$ with the
aim of mapping the interacting trajectories into non interacting
ones of an effective bath.

For values of densities ranging from $0$ (non interacting) to $0.7$, we are able to describe the active system
through a set of effective parameters: tumbling rate $\lambda_{eff}$, velocity
of  bacterium in the ballistic regime $v_{eff}$ and time spent in tumbling state $\tau_{eff}$. 
All these density dependent parameters display an additional dependence on the wave vector $q$. In particular, the effective swimming speed is an increasing function
of $q$ that tends to the single particle run speed at large $q$ values.
Recent DDM experiments on wild type, tumbling {\it E.coli} \cite{Wilson11} have shown a similar q dependence in the swimming speed when fitting experimental data with a non tumbling model for the ISF. In that situation, the smaller speed observed at lower wave vectors was explained as the effect of tumbling that breaks straight runs and results on average speeds that are smaller when observed over longer length/time scales. Here a similar effect is observed when fitting data to a full model that includes tumbling but no interactions.
The presence of bacterial collisions produces an increased deviation of trajectories from straight runs  and is also responsible for the $q$ dependent reduced speed. Collisions are also responsible for an higher effective tumbling rate that is found to be a decreasing function of $q$.
For all $q$ values, the effective time spent in a tumbling state goes to zero when 
density increases. 
Our results could also suggest new methodologies for retrieving accurate motility characteristics from experimental dynamic structure factors of concentrated bacterial suspensions.

\ack
We acknowledge support from MIUR-FIRB project RBFR08WDBE and CASPUR
High Performance Computing initiatives. The research leading to these results has received funding from the European Research Council under the European Union's Seventh Framework Programme (FP7/2007-2013) / ERC grant agreement n$^\circ$ 307940.
\appendix
\section{Computation of ISF}
Through the effective motility parameters we can compute the ISF shown in right panel of Fig. (\ref{fig:sqwfqt}).
Eqs. (\ref{luca}) define the exact propagator for run-and-tumble dynamics. In order
to compute the ISF we have to invert such equations in time-Fourier space $({\bf q},t)$.
It is known that, for an instantaneous run-and-tumble dynamics ($\tau=0$), the exact
propagator can be written as a power series in $\lambda$\footnote{In internal unit $\lambda=0.1$.} \cite{Martens12}
\BEQ
\mathscr{F}({\bf q},z)=\frac{F_{0}({\bf q}, z)}{1- \lambda F_{0}({\bf q}, z)}=  \sum_{n=0}^\infty \lambda^n  F_{0}({\bf q}, z )^{n+1}\, .
\EEQ
Than function $\mathscr{F}({\bf q},z,\tau)$ reads  
\BEA\label{pert}
\mathscr{F}({\bf q},z,\tau) &=& \sum_{n=0}^\infty \lambda^n  F_{0,n+1}({\bf q}, z , \tau) \\ \nn
F_{0,n}({\bf q}, z , \tau) &\equiv& e^{- (n -1)z \tau} F_0^n({\bf q},z) 
\EEA
where $F_0({\bf q},z)$ is the ballistic propagator.
In time-Fourier space perturbative expansion is
%
%
%\begin{widetext}
\BEA \nn 
\mathscr{F}({\bf q},t,\tau) &=& F_{0,1}({\bf q}, t , 0) + \lambda \int_{0}^{t } \, dt^\prime \, F_{0,1}({\bf q}, t-t^\prime , 0) F_{0,1}({\bf q}, t^\prime , \tau) \\ \nn
 &+& \lambda^2 \int_{0}^{t } \, dt^\prime \, F_{0,1}({\bf q}, t-t^\prime , 0)F_{0,2}({\bf q}, t^\prime , \tau) + o(\lambda^3) \\ \label{exp} 
F_{0,n}({\bf q}, t , \tau) &\equiv&\mathcal{L}^{-1}\left[ e^{- n z \tau} F_0^n({\bf q},z) \right] \, .
\EEA
 %\end{widetext}
%
%
%
From Eq. (\ref{inv}), using Eq. (\ref{exp}) follows 
%\begin{widetext}
\BEA \label{sol}
P(q,t,\tau)&=&\frac{1}{1 + \lambda \tau}\left[ \sum_{n=0}^\infty \lambda^n F_{0,n+1}(t - n \tau)\theta(t - n \tau) \right. \\ \nn
 &+&\sum_{n=0}^\infty\left(I_{1,n} + I_{2,n} + I_{3,n} \right) \\ \nn &+&
\left. \lambda \left( t - \tau \right) \left( \theta(t - \tau) - 1 \right) \vphantom{\sum_{n=0}^{\inf}} \right] \\ \nn
I_{1,n}&\equiv& \lambda^{n + 1} \int_{0}^{t - n \tau} dt^\prime \, F_0^{n+1} (t - t^\prime - \tau) \left( 2 + \lambda t^\prime \right)  \\   \nn 
I_{2,n}&\equiv& \lambda^{n + 2} \int_{0}^{t - (n+2) \tau} dt^\prime \, F_0^{n+1} (t - t^\prime - (n+2) \tau) t^\prime  \\ \nn 
I_{3,n}&\equiv& -2 \lambda^{n + 1}\int_{0}^{t - (n+1) \tau} dt^\prime \, F_0^{n+1} (t - t^\prime - (n+1) \tau) \left( 1 + \lambda t^\prime\right) \, .
\EEA
%\end{widetext}
%
In two dimensions, the anti-Laplace transform of an arbitrary
power of $F_0({\bf q},z)$ is\cite{Martens12}
\BEQ\label{pote}
F_{0,n+1}(q, t) =\frac{e^{-\lambda t} \sqrt{\pi}}{2^{\frac{n}{2}}\Gamma\left( \frac{n+1}{2}\right)}\left( \frac{t}{q v} \right)^{\frac{n}{2}}J_{\frac{n}{2}}(q v t) \, ,
\EEQ
where $J_m$ is the Bessel function of the first kind, $\Gamma(x)$ is the Euler gamma.
Putting Eq. (\ref{pote}) in Eq. (\ref{sol}), through a numerically integration, we can compute ISF
in any order $n$.


\section*{References}
\begin{thebibliography}{99}
%\bibitem{Vicsek12} T. Vicsek and A. Zafeiris, Phys. Rep. {\bf{517}}, 71 (2012). 
%\bibitem{Najafi04} A. Najafi, and R. Golestanian, Phys. Rev. E {\bf{69}}, 062901 (2004).
%\bibitem{Golestanian08} R. Golestanian, and A. Ajdari, Phys. Rev. E {\bf{77}}, 036308 (2008).
%\bibitem{Pooley07} C. M. Pooley, G. P. Alexander, and J. M. Yeomans, Phys. Rev. Lett. {\bf{99}}, 228103 (2007).
%\bibitem{DiLeonardo11} R. Di Leonardo, D. Dell' Arciprete, L. Angelani, and V. Iebba, Phys. Rev. Lett {\bf{106}}, 038101(2011).
\bibitem{Vicsek95} T. Vicsek, A. Czirok, E. Ben-Jacob, I. Cohen, and O. Shochet, Phys. Rev. Lett. {\bf{75}}, 1226 (1995).
\bibitem{Chate08} H. Chat\'e, F. Ginelli, G. Gr\'egoire, F. Peruani, and F. Raynaud, Eur. Phys. J. B {\bf{74}}, 451 (2008). 
\bibitem{Chate08b} H. Chat\'e, F. Ginelli, G. Gr\'egoire, and F. Raynaud, Phys. Rev. E {\bf{77}}, 046113 (2008). 
\bibitem{Baskaran08} A. Baskaran, and M. C. Marchetti, Phys. Rev. Lett. {\bf{101}}, 268101 (2008).
\bibitem{Baskaran09} A. Baskaran, and M. C. Marchetti, Proc. Natl. Acad. Sci. U.S.A. {\bf{106}}, 15567 (2008).
\bibitem{Bertin06} E. Bertin, M. Droz, and G. Gr\'egoire, Phys. Rev. E {\bf{74}}, 022101 (2006).
\bibitem{Bertin09} E. Bertin, M. Droz, and G. Gr\'egoire, J. Phys. A: Math. Theor. {\bf{42}}, 445001 (2009).
%
\bibitem{Farrell12} F. D. C. Farrell, M. C. Marchetti, D. Marenduzzo, and J. Tailleur, Phys. Rev. Lett. {\bf{108}}, 248101 (2012).
%
\bibitem{Csahok02} Z. Csah\'ok, and A. Czir\'ok, Physica A {\bf{243}} 304, (2002).
\bibitem{Toner95} J. Toner, and T. Tu Phys. Rev. Lett. {\bf{75}} 4326, (1995).
\bibitem{Toner98} J. Toner, and T. Tu Phys. Rev. E {\bf{58}} 4828, (1995).
%
\bibitem{Schnitzer93} M. J. Schnitzer, Phys. Rev. E {\bf{48}}, 2553 (1993).
\bibitem{Cates12} M. E. Cates, Rep. Prog. Phys. {\bf{75}}, 042601, (2012).
\bibitem{Berg04} H. C. Berg, {\itshape E. Coli In Motion} (Springer, New York, 2004).
\bibitem{Berg72} H. C. Berg, D. A. Brown, Nature (London) {\bf{239}}, 500 (1972).
\bibitem{Grossart01} H. -P. Grossart, L. Riemann, and F. Azam, Aquatic Microbial Ecology {\bf{25}}, 247 (2001).
\bibitem{Schneider74} W. R. Schnieder, and R. N. Doetsch, J. Bacterial. {\bf{117}}, 696 (1974). 
\bibitem{Othmer88} H. G. Othmer, S. R. Dunbar, W. Alt, J. Mat. Biol. {\bf{26}}, 263 (1988).
\bibitem{Martens12} K. Martens, L. Angelani, R. Di Leonardo and L. Bocquet, Eur. Phys. J. E {\bf{35}}, 84 (2012).
\bibitem{Wilson11} L. G. Wilson, V. A. Martinez, J. Schwartz-Linek, J. Tailleur, G. Bryant, P. N. Pusey and W. C. K. Poon, Phys. Rev. Lett. {\bf{106}}, 018101 (2011).
\bibitem{Martinez12} V. A. Martinez, R. Besseling, O. A. Croze, J. Tailleur, M. Reufer, J. Schwartz-Linek, L. G. Wilson, M. A. Bees, and W. C. K. Poon, Biophys. J. {\bf{103}}, 1637 (2012).
\bibitem{Cerbino08} R. Cerbino, and V. Trappe, Phys. Rev. Lett. {\bf{100}}, 188102 (2008).
\bibitem{Boon74} J. P. Boon, R. Nossal, and S. H. Chien, Biophys. J. {\bf{14}}, 847 (1974).
\bibitem{Stock78} G. B. Stock, Biophys. J. {\bf{22}}, 79, (1978).
\bibitem{Giavazzi09} F. Giavazzi, D. Brogioli, V. Trappe, T. Bellini, and R. Cerbino, Phys. Rev. E {\bf{80}}, 031403 (2009)
\bibitem{Angelani13} L. Angelani,  EPL {\bf{102}}, 20004 (2013). 
%\bibitem{Davis84} M. H. A. Davis, J. R. Statist. Soc. B {\bf{46}}, No. 3, 353, (1984).
\bibitem{Tailleur08} J. Tailleur, and M. E. Cates, Phys. Rev. Lett. {\bf{100}}, 218103 (2008).
\bibitem{Angelani09} L. Angelani, R. Di Leonardo and G. Ruocco, Phys. Rev. Lett. {\bf{102}}, 048104 (2009).
\bibitem{Angelani10} L. Angelani and R. Di Leonardo, New Journal of Physics {\bf{12}}, 113017 (2010).
\bibitem{Angelani11} L. Angelani and R. Di Leonardo, Comp. Phys. Commun. {\bf{182}}, 1970 (2011).
%\bibitem{Angelani11b} L. Angelani, A. Costanzo and R. Di Leonardo, EPL {\bf{96}}, 68002 (2011).
%
%
%\bibitem{Cates10} M. E. Cates, D. Marenduzzo, I. Paganobarraga, and J. Tailleur, Proc. Natl. Acad. Sci. U. S. A. {\bf{107}}, 11715 (2010).
\bibitem{Purcell77} E. M. Purcell, Am. J. Phys. {\bf{45}}, 3 (1977).
\bibitem{Kim05} S. Kim and S. Karrila, {\itshape Microhydrodynamics} (Dover, New York, 2005).
\bibitem{Press92} W. H. Press, W. T. Wetterling, S. A. Teukolsky, B. P. Flannery, Numerical Recipes in C, Cambridge University Press, Cambridge, (1992).
\bibitem{Klafter11} Klafter J. and Sokolov I. M., First Steps in Random Walks (Oxford University Press, New York.) 2011
%\bibitem{Jepson13} A. Jepson, V. A. Martinez, J. Schwarz-Linek, A. Morozov, W. C. K. Poon, http://arxiv.org/abs/1307.1274 (2013).
\bibitem{Cates13} M. E. Cates and J. Tailleur EPL {\bf{101}}, 20010 (2013).
%\bibitem{Boon74} J.P. Boon, R. Nossal, and S.H. Chien Biophys. J., {\bf{14}} (1974).
\bibitem{Berne00} B.J. Berne, R. Pecora Dynamic Light Scattering Dover, Mineola, NY (2000).
 %{\bf{4}} (1971) 1071.
%
%
%
\end{thebibliography}
\end{document}